\documentclass[12pt]{article}

\newenvironment{sciabstract}{%
\begin{quote} \bf}
{\end{quote}}

\newcounter{lastnote}

\title{A simple {\it Science\/} Template}

\author
{John Smith,$^{1}$ Jane Doe,$^{1}$ Joe Scientist$^{2\ast}$\\
\\
\normalsize{$^{1}$Department of Chemistry, University of Wherever,}\\
\normalsize{An Unknown Address, Wherever, ST 00000, USA}\\
\normalsize{$^{2}$Another Unknown Address, Palookaville, ST 99999, USA}\\
\\
\normalsize{$^\ast$To whom correspondence should be addressed; E-mail:  jsmith@wherever.edu.}
}

\date{}

\usepackage[pdftex]{graphicx}
\usepackage{subfigure}
\usepackage{ams math, amsfonts}
\usepackage{url}
\usepackage[usenames, dvipsnames]{color}
\newcommand{\CUT}[1]{}

\newcommand{\mtnoter}[1]{}
\DeclareMathOperator*{\Tr}{Tr}
\DeclareMathOperator*{\diag}{diag}

\title{A Single Model Explains both Visual and Auditory Precortical Coding}

\author{Honghao Shan,$^{1}$ Matthew H.~Tong,$^{1}$ Garrison W.~Cottrell$^{1\ast}$\\
\\
\normalsize{$^{1}$Department of Computer Science and Engineering, University of California, San Diego}\\
\normalsize{La Jolla, CA 92093--0404}\\
\\
\normalsize{$^\ast$To whom correspondence should be addressed; E-mail: gary@ucsd.edu.}
}

\begin{document}




\maketitle

\begin{sciabstract}
Precortical neural systems encode information collected by the senses, but the driving principles of the encoding used have remained a subject of debate. We present a model of retinal coding that is based on three constraints: information preservation, minimization of the neural wiring, and response equalization. The resulting novel version of sparse principal components analysis successfully captures a number of known characteristics of the retinal coding system, such as center-surround receptive fields, color opponency channels, and spatiotemporal responses that correspond to magnocellular and parvocellular pathways. Furthermore, when trained on auditory data, the same model learns receptive fields well fit by gammatone filters, commonly used to model precortical auditory coding. This suggests that efficient coding may be a unifying principle of precortical encoding across modalities.
\end{sciabstract}

\section*{Introduction} \label{sec:spca:introduction}

Sensory information goes through various forms of processing before it reaches the cerebral cortex. Visual information is transformed into neural signals at the retina, where it passes through retinal ganglion cells that are characterized by their center-surround shaped receptive fields~\cite{Cugell1966RetinaGanglionCells}; auditory information, on the other hand, is passed to the brain through the auditory nerve fibers whose filtering properties can be well described by gammatone filters~\cite{Kiang1965AuditoryNerveFibers}. Since such peripheral processing prepares the data that the subsequent cortical processing relies on, its functional role has attracted a great deal of attention in the past several decades~\cite{Laughlin1982RetinaWhitening,Field1987Whitening,Atick1992RetinaDecorrelation,Lewicki2002NaturalSounds,Vincent2005Retina,Field2006WhiteningTheory,Doi2007RetinalCoding}.

Despite intensive research, there are still mysteries concerning the functional role of pre-cortical processing. For example, do different sensory modalities (visual, auditory, somatosensory, etc.) adopt the same computational principles in their pre-cortical stages? Although it is tantalizing to assume so, recent studies suggest otherwise. For example, \cite{Lewicki2002NaturalSounds} learned gammatone filters from natural sound using independent component analysis (ICA), which was previously applied to natural image patches to learn edge/bar shaped filters resembling the V1 simple cells' receptive fields~\cite{Olshausen1996SparseCoding_SimpleCell,Lewicki2000ICA_Overcomplete}. Since gammatones model pre-cortical auditory nerve fibers while V1 is a region of cortex, this gives rise to a puzzle: Why would the brain use the same strategy for preprocessing at a \emph{pre-cortical} stage in the auditory pathway and early \emph{cortical} processing in the visual pathway~\cite{Olshausen2002NewWindow}?

Questions remain even for peripheral processing in a single modality. Recently, Graham et al. proposed that decorrelation, response equalization, and sparseness form the minimum constraints that must be considered to account for the known linear properties of retinal coding \cite{Field2006WhiteningTheory}. This hypothesis is the combination of several previous theories. The response equalization theory hypothesizes that retinal coding seeks a representation that ensures that each neuron has approximately the same average activity level when the animal is presented with natural scenes~\cite{Field1987Whitening,Field1995whitening,Field1997Whitening,Field2000whitening}. 
The output decorrelation theory follows the efficient coding principle~\cite{Attneave1954SparseCoding_Redundancy,Barlow1961SparseCoding,Atick1992RetinaDecorrelation}. It hypothesizes that retinal coding represents the most efficient coding of the information in the visual domain by capturing the second-order statistical structure of the visual inputs and making the signals from these neurons less correlated. Both of these theories are derivatives of whitening theory, which hypothesizes that retina coding produces a flattened response spectrum for natural visual inputs from a specific range of spatial frequencies~\cite{Laughlin1982RetinaWhitening}. This whitening theory links the properties of retinal coding with the statistics of natural scenes and is now part of the prevailing view of retinal processing. A third theory suggests that the system is trying to minimize energy usage or wiring cost~\cite{Vincent2003EnergyRetina}. Vincent et al. argued that systems that try to minimize energy usage by minimizing wiring give center surround receptive fields.  Given these various objectives, it is still not clear what constraints are actually operating in the specification of the retinal coding system.
It would be desirable to build a retinal coding model that integrates the different ideas behind these theories and explains the origins of the observed center-surround receptive fields. Ideally, this model should also able to explain the pre-cortical processing of other modalities.

\begin{figure}
    \begin{center}
    \subfigure[PCA Network]{\includegraphics[height=0.20\textwidth]{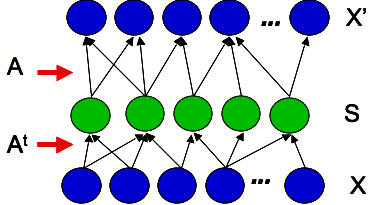}\label{fig:PCA-network}}
    \subfigure[Sparse PCA Network]{\includegraphics[height=0.20\textwidth]{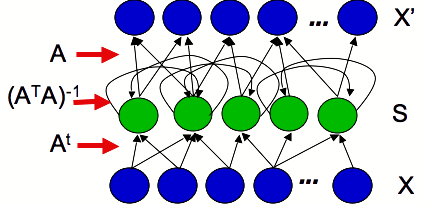}\label{fig:SPCA-network}}
    \end{center}
    \caption{Two neural networks that can implement PCA and Sparse PCA. The left hand panel represents a network that performs PCA. The weights (rows of $\mathbf{A^t}$) to each hidden unit (ganglion cell) from the pixels (photoreceptors) represent the coordinates of a unit-length eigenvector of the covariance matrix of the data, so that the activations are the projections onto the eigenvectors. The same weights (transposed) can be used to reconstruct the data (these are not part of the model). The right panel represents Sparse PCA. It is the connections from the pixels that are sparse ($\mathbf{A^t}$). This is followed by recurrent connections that give the center-surround shape. This would be reflected in any recordings of the hidden units, hence the receptive field of a hidden unit is represented by a row of $\mathbf{W=(A^tA)^{-1}A^t}$. 
        \label{fig:networks}}
\end{figure}

Our model takes into account the following considerations. The retina compresses the approximately 100 million photoreceptor responses into a million ganglion cell responses. Hence the first consideration is that we would like the ganglion cells to retain the maximum amount of information about the photoreceptor responses. If we make the simplifying assumption that ganglion cells respond linearly, then the optimal linear compression technique in terms of reconstruction error is principal components analysis (PCA). One can map PCA into a neural network as in Figure ~\ref{fig:PCA-network}~\cite{cmz:sharkey,Baldi89}. The weight vectors of each hidden unit in this network each correspond to one eigenvector of the covariance matrix of the data.  In standard PCA, there is an ordering to the hidden units, such that the first hidden unit has very high response variance and the last hidden unit has practically no variance, which means the first hidden unit is doing orders of magnitude more work than the last one. The second consideration, then, is that we would like to spread the work evenly among the hidden units. Hence we impose a threshold on the average squared output of the hidden units. As we will see from the simulations, in order to preserve the maximum information, the units all hit this threshold, which equalizes the work. The third consideration is that PCA is profligate with connections - every ganglion cell would have non-zero connections to every photoreceptor. Hence we also impose a constraint on the connectivity in the network. In this latter constraint we were inspired by the earlier work of \cite{Vincent2005Retina}. They proposed a model of retinal and early cortical processing based on energy minimization and showed that it could create center-surround shaped receptive fields for grayscale images.  However, their system sometimes led to cells with two center-surround fields, and the optimization itself was unstable. 

These considerations lead to our objective function. Images can be represented as high-dimensional real-valued data; if L photoreceptors are representing the input image, the observed image can be represented as $x \in R^L$. Given input vectors $\mathbf{x} \in R^L$, we  seek to find the output responses $\mathbf{s} \in R^M$ (the signal from the retinal ganglion cells) and basis functions $A \in R^{L \times M}$  (the connections from the photoreceptors to the ganglion cells) such that the following objective function is minimized:

\begin{eqnarray}
E = \left< \frac{\|\mathbf{x} - \mathbf{A}\mathbf{s}\|_2^2}{2} \right> + \lambda \|\mathbf{A}\|_1 \label{eq:obj_fn_1}
\end{eqnarray}
subject to:
\begin{eqnarray}
\left<s_i^2\right> \leq 1 ~~\forall i \label{eq:obj_fn_2}
\end{eqnarray}
where $\left< \cdot \right>$ denotes taking average over all the input samples. 
The first term in Equation \ref{eq:obj_fn_1} minimizes the reconstruction error and maximizes the information maintained by the encoding. When the sparsity weight $\lambda$ is small and $L>M$ (i.e., the encoding compresses the information), the reconstruction error reduces to a term that only involves the correlation matrix: $\mathbf{C} = \left<\mathbf{x}\mathbf{x}^t \right>$ (see supplementary materials). This concurs with the idea that the system is only sensitive to second order statistics.  The second constraint minimizes the connections from the photoreceptors to the ganglion cells, incorporating sparsity and an economy of elementary features. The constraint on the average energy of the ganglion cells equalizes the work across the ganglion cells. The system will in fact push this term to the threshold of 1 in order to maintain the maximum information. Thus this objective function integrates three major theories of retinal coding: efficient coding~\cite{Attneave1954SparseCoding_Redundancy}, response equalization~\cite{Field1987Whitening}, and the economy of elementary features~\cite{Vincent2005Retina}. While this does not directly embody the decorrelation theory~\cite{Atick1992RetinaDecorrelation},  we have not found that assumption necessary to obtain our results. 

Put another way, the three terms in the objective function determine different aspects of the basis functions (i.e., the columns of $\mathbf{A}$): the reconstruction error determines the subspace that the basis functions span; the output constraint specifies the lengths of the basis functions; and the sparsity penalty rotates the basis functions within the subspace determined by the reconstruction error. In this sense, all three terms, and hence all three theories that they embody, are necessary to fully characterize the retinal coding model. This observation partially agrees with the prediction by \cite{Field2006WhiteningTheory}: ``we conclude that a minimum of three constraints must be considered to account for the known linear properties -- decorrelation, response equalization, and size/sparseness.''  Our three constraints are efficient coding of information, response equalization, and sparseness of connections.

An important feature of this objective function is that it allows us to derive an efficient algorithm to learn the model parameters, because the revised model turns out to be a particular variation of Sparse PCA~\cite{Tibshirani2006SPCA} that is reducible to sparse coding~\cite{Olshausen1996SparseCoding_SimpleCell}. We can therefore efficiently estimate the parameters of the model and apply it to a larger range of data than has typically been used in the past.
A critical insight provided by the mapping to sparse coding is that this model  {\em is} exactly sparse coding, applied to the {\em transpose} of the data matrix rather than the data matrix itself. This means we can also use our model for dimensionality {\em expansion} (overcomplete representations) as well as dimensionality reduction, although when doing expansion, we can no longer use the efficient approximation derived in the Supplementary Materials.

In what follows, we show that this simple objective function is able to account for both retinal ganglion cell receptive fields and gammatone filters that have been used to characterize the signals in the auditory nerve.

\section*{Results}

Going forward, it is important to understand the distinction between {\em features} and {\em filters}. The features are the the rows of $\mathbf{A^t}$, and represent the connections between the photoreceptors and the hidden units in Figure~\ref{fig:SPCA-network}. The filters, on the other hand, correspond to the rows of $\mathbf{W}=\mathbf{(A^tA)^{-1}A^t}$, the pseudoinverse of $\mathbf{A}$, which correspond to the receptive fields of the  ganglion cells that would result from reverse correlation. We visualize the computation as in Figure~\ref{fig:SPCA-network}: the network receives input from the photoreceptors, and then there is inhibition of hidden units with similar receptive fields (represented by the recurrent connections $\mathbf{(A^tA)^{-1}}$).

\paragraph*{Grayscale Images} \label{sec:spca:experiments:grayscale}

\begin{figure}
    \begin{center}
    \subfigure[Distribution of connection weights using SPCA]{\includegraphics[width=0.49\textwidth]{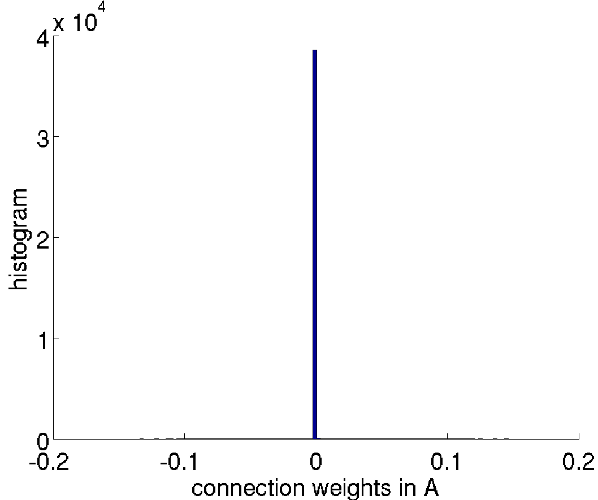}\label{fig:spca:grayscale:method:spca}}
    \subfigure[Distribution of connection weights using PCA]{\includegraphics[width=0.49\textwidth]{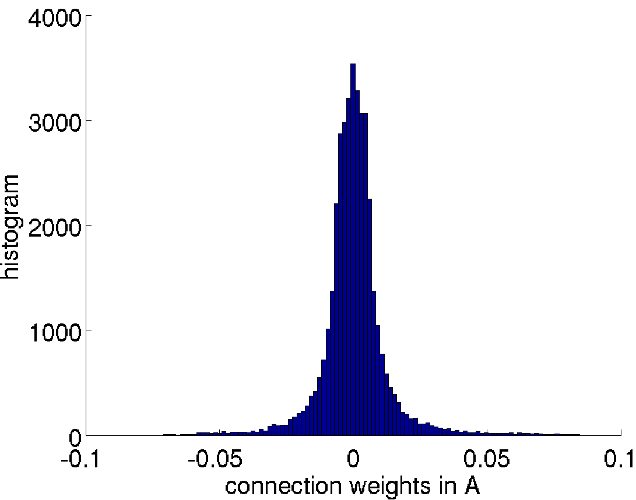}\label{fig:spca:grayscale:method:pca}}
    \end{center}
    \caption{Comparison of SPCA and PCA connection weights on grayscale image patches.  Figure~\ref{fig:spca:grayscale:method:spca} and \ref{fig:spca:grayscale:method:pca} plot the distribution of the connection weights in $\mathbf{A}$ learned by our model versus those learned by standard PCA. }
      \label{fig:spca:grayscale:method}
\end{figure}

We applied the model to four grayscale image datasets. Results were qualitatively similar across the sets; the results we describe here are from the subset of the Van Hateren natural image set described in  \cite{Lewicki2009ComplexCells}. For this simulation, we used 20$\times$20 patches of pixels and reduced them to 100 dimensions. 
Our SPCA model captures $99.23\%$ of the variance that is captured by standard PCA with 100 eigenvectors retained, while $96.31\%$ of the connection weights (the rows of $\mathbf{A^t}$) are absolute zero; in contrast, none of the connection weights in standard PCA are zero. Figure~\ref{fig:spca:grayscale:method:spca} and \ref{fig:spca:grayscale:method:pca} plots the distribution of the connection weights in $\mathbf{A}$ learned by our model versus standard PCA. 
Each ganglion cell is directly connected to only $3.69\%$ of the input neurons in the 20$\times$20  patch on average. Clearly, this sparsity would be advantageous for a biological system. 

\begin{figure}
    \begin{center}
    \subfigure[Elementary features and filters]{\includegraphics[width=0.4\textwidth]{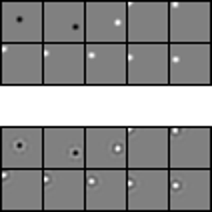}\label{fig:spca:grayscale:features:AW}}
    \subfigure[Distribution of $100$ elementary features]{\includegraphics[width=0.4\textwidth]{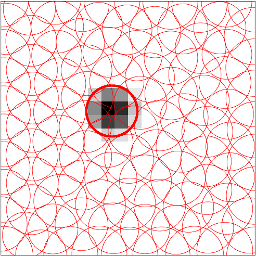}\label{fig:spca:grayscale:features:fit100}}\\
    \subfigure[Filter as a weighted sum of features]{\includegraphics[width=0.4\textwidth]{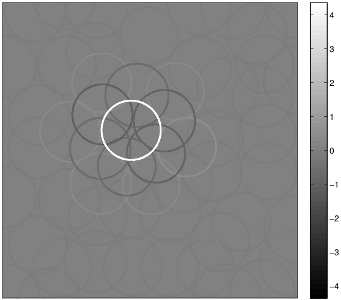}\label{fig:spca:grayscale:features:cs100}}    
    \subfigure[Distribution of $32$ elementary features]{\includegraphics[width=0.4\textwidth]{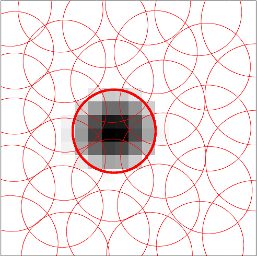}\label{fig:spca:grayscale:features:fit32}}
    \end{center}
    \caption{Elementary features and their corresponding optimal filters, learned from $20 \times 20$ grayscale image patches. In Figure~\ref{fig:spca:grayscale:features:AW}, the top panel displays $10$ features randomly selected from all the $100$ features; the bottom panel displays their optimal filters. We fit the features with Gaussian blobs and plot them as circles in Figure~\ref{fig:spca:grayscale:features:fit100}. The radius of each circle represents twice the standard deviation of the fitted Gaussian blob. To help visualize how well the Gaussian blobs fit with the features, we display the first feature in Figure~\ref{fig:spca:grayscale:features:AW} and highlight its fitted Gaussian. These Gaussian will become bigger if we use a smaller number of features to ``construct'' the image patches, as shown in Figure~\ref{fig:spca:grayscale:features:fit32}. Figure~\ref{fig:spca:grayscale:features:cs100} displays the first filter in Figure~\ref{fig:spca:grayscale:features:AW} as a weighted sum of all the $100$ features. Each feature is plotted as a circle, as in Figure~\ref{fig:spca:grayscale:features:fit100} and \ref{fig:spca:grayscale:features:fit32}. The color of each circle represents the weight assigned to this feature (read the main text for details).}
    \label{fig:spca:grayscale:features}
\end{figure}

The learned elementary features (i.e., the columns of $\mathbf{A}$) are blobs of similar size that tile the $20 \times 20$ image patch. The top panel in Figure~\ref{fig:spca:grayscale:features:AW} displays $10$ features randomly selected from all $100$ features. We fit all the features with 2D Gaussians, and plot them as circles in Figure~\ref{fig:spca:grayscale:features:fit100}. The center and the radius of each circle represent the center and twice the standard deviation of the fitted Gaussian. To visualize how well the Gaussians fit the features, we display the first feature in Figure~\ref{fig:spca:grayscale:features:AW} and highlight its fitted Gaussian. As shown in the figure, the Gaussians provide a mosaic coverage of the image patch. If we reduce the number of hidden units to 32, the blobs enlarge to cover the image, as shown in  Figure~\ref{fig:spca:grayscale:features:fit32}.

The optimal filters (i.e., the rows of $\mathbf{W}$) are center-surround shaped, as shown in the bottom panel of Figure~\ref{fig:spca:grayscale:features:AW}. The first filter, for example, recovers the weight assigned to the first feature in the top panel. It is tantalizing to think that some of the filters are ON-centered while others are OFF-centered. However, we can switch the signs of the features (and hence the signs of the optimal filters) without changing the model's objective function. Hence our model does not provide insight into the difference between the ON-centered and the OFF-centered cells~\cite{Chichilnisky2002OnOffRGC} beyond the usual explanation that neurons cannot fire both positively and negatively.

It is interesting to see why a population of Gaussian blob shaped features should give rise to center-surround shaped filters.  As shown in Figure~\ref{fig:spca:grayscale:features:cs100}, each filter is a weighted sum of all the elementary features and can be viewed as the result of a sequence of efforts to recover the contribution of its corresponding feature. Each feature is first applied as a template filter on the image patch to estimate its contribution. However, this estimation is inaccurate because this feature overlaps with its neighboring features. To get a more accurate estimation the contribution from neighboring features must be subtracted. This potentially overcompensates, so get an even more accurate estimation, one must add back the contribution from the features neighboring the features that surround the first feature. This process repeats, moving ever outward. However, the weight reduces quickly for features removed from the first feature, which makes the resulting filter effectively localized and keeps the filter center-surround shaped (for low lambda, some additional ripples can be observed - however, these additional ripples have also been observed in ganglion receptive fields \cite{Dearworth2002RGC}).

\begin{figure}
    \begin{center}
    \subfigure[White noise]{\includegraphics[width=0.3\textwidth]{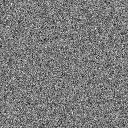}\label{fig:spca:grayscale:noise:white}}
    \subfigure[Pink noise]{\includegraphics[width=0.3\textwidth]{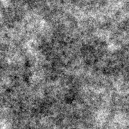}\label{fig:spca:grayscale:noise:pink}}
    \subfigure[Learned filters]{\includegraphics[width=0.3\textwidth]{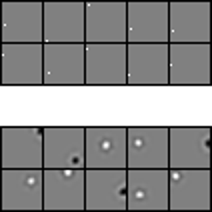}\label{fig:spca:grayscale:noise:W}}
    \end{center}
    \caption{Experiments on white noise and pink noise images. Figure~\ref{fig:spca:grayscale:noise:white} and \ref{fig:spca:grayscale:noise:pink} display images containing white noise and pink noise. The top panel in Figure~\ref{fig:spca:grayscale:noise:W} displays $10$ filters learned from white noise images; the bottom panel plays the filters learned from pink noise images.}
    \label{fig:spca:grayscale:noise}
\end{figure}

Which aspects of natural scene images give rise to the learned features we observe? To answer this question, we apply our algorithm to white noise images, which contain no statistical structure, and pink noise images, which follow the same $1/f$ power law as natural scene images~\cite{Field1987Whitening} but otherwise contain no structure. Figure~\ref{fig:spca:grayscale:noise} displays two example images and some of the learned filters. On white noise images, the learned features  are one-pixel image templates; the corresponding filters also only contain one non-zero pixel. That is, the model simply keeps 64 pixels and ignores the other pixels. That's the best it can do with $64$ features, because white noise images contain no structure. On pink noise images, we learn essentially the same elementary features (and hence the same filters) as those learned from natural scene images. This result supports the hypothesis that the center-surround shaped filters come from the $1/f$ power spectrum of natural scene images, which agrees with the classic whitening theory~\cite{Laughlin1982RetinaWhitening,Field1987Whitening,Atick1992RetinaDecorrelation}.

The sparseness level $\lambda$ plays an important role in shaping the learned features and filters. As $\lambda$ increases, the model puts more emphasis on sparse connections at the cost of keeping less information about the inputs. In a biological system, this may occur when the system is on a strict energy budget. Here we check how the learned filters change with larger $\lambda$ values.

\begin{figure}
    \begin{center}
    \includegraphics[width=0.6\textwidth]{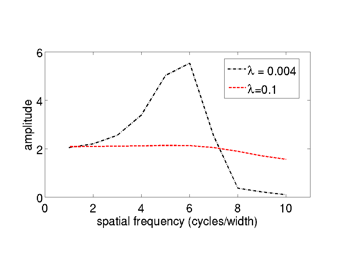}\label{fig:spca:grayscale:sparseness:sensitivity}
    \end{center}
    \caption{Experiments with increased sparseness level $\lambda$. We plot the amplitudes of different frequency component of the filters learned with $\lambda=0.004$ and $\lambda=0.1$. As shown in the figure, with an increased $\lambda$ value, the filter becomes less sensitive to low frequencies, but more sensitive to high frequencies.}
    \label{fig:spca:grayscale:sparseness}
\end{figure}

By analyzing the filters in Fourier space, we can plot amplitude at various frequencies, giving a contrast sensitivity function. 
As shown in Figure~\ref{fig:spca:grayscale:sparseness}, with larger $\lambda$ value, the model becomes less sensitive to low frequency information, but more sensitive to high frequency information.  This change matches with psychophysical studies of contrast sensitivity in children with chronic malnutrition. Compared with normal children, malnourished children are reported to be less sensitive to low spatial frequencies, but slightly more sensitive to high spatial frequencies~\cite{dosSantos2010MalnutritionedKids}. This shift of acuity towards high frequencies, as suggested by our result, might due to the effort of the neural system to capture more visual information with a limited neural wiring budget. 

\paragraph*{Chromatic Images} \label{sec:spca:experiments:color}

We applied our algorithm to four chromatic image datasets and again found that we learn qualitatively similar features with each. Here, we report the features learned from Kyoto image dataset. Retinal L, M, and S cones are estimated and given as input to the model. The resulting model captures $99.75\%$ of the variance that is captured by an optimal linear model (PCA) with $256$ output neurons, with $96.11\%$ of its connections being absolute zero.

\begin{figure}
    \begin{center}
    \subfigure[Elementary features and filters]{\includegraphics[width=0.455\textwidth]{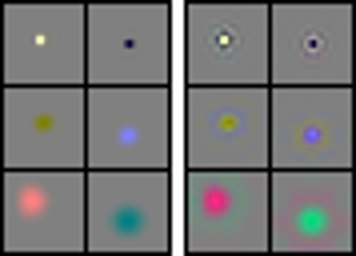}\label{fig:spca:color:feature:AW}}
    \subfigure[Connection from the filters to the LMS cones]{\includegraphics[width=0.45\textwidth]{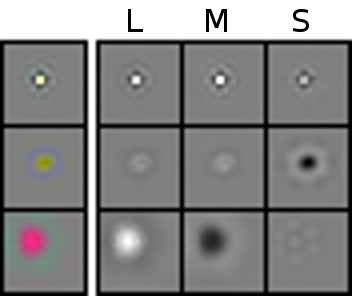}\label{fig:spca:color:feature:W3}}\\
    \subfigure[Red/green]{\includegraphics[width=0.25\textwidth]{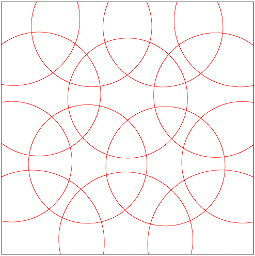}\label{fig:spca:color:feature:red}}
    \subfigure[Blue/yellow]{\includegraphics[width=0.25\textwidth]{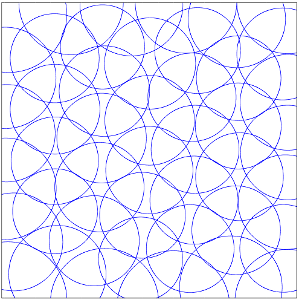}\label{fig:spca:color:feature:blue}}
    \subfigure[Black/white]{\includegraphics[width=0.25\textwidth]{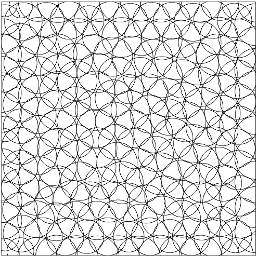}\label{fig:spca:color:feature:black}}
    \end{center}
    \caption{Elementary features and the corresponding filters learned from $20 \times 20$ chromatic image patches with $\lambda=0.001$. In Figure~\ref{fig:spca:color:feature:AW}, the left panel displays $6$ representative features; the right panel displays their corresponding filters. The features belong to three categories: black/white blobs, blue/yellow blobs, and red/green blobs. The corresponding optimal filters are center-surround shaped, with black/white, blue/yellow, or red/green antagonism. Figure~\ref{fig:spca:color:feature:W3} plots the connection strength from the filters to the L, M, S cones. Figure~\ref{fig:spca:color:feature:red}, \ref{fig:spca:color:feature:blue}, and \ref{fig:spca:color:feature:black} plot the spatial layout of learned features, as we did in Figure~\ref{fig:spca:grayscale:features:fit100}.}
    \label{fig:spca:color:feature}
\end{figure}

Figure~\ref{fig:spca:color:feature:AW} displays $6$ representative features as well as their corresponding filters, learned from chromatic image patches. We visualize the connection strength from three of the filters to the L/M/S channels in Figure~\ref{fig:spca:color:feature:W3}. Among all the $256$ learned features, $193$ are black/white blobs, $48$ are blue/yellow blobs, $15$ are red/green blobs. Figure~\ref{fig:spca:color:feature:red}, \ref{fig:spca:color:feature:blue}, and \ref{fig:spca:color:feature:black} plot the spatial layout of learned features.

The above result replicates the segregation of the spatial channel and the color channel at the retina stage~\cite{Calkins1999SpaceColorsegregation}. This segregation was explored in previous research that applied information-theoretic methods to natural color spectra/images, such as PCA~\cite{Buchsbaum1983Trichromacy,Buchsbaum1991ColorPCA} and ICA~\cite{Buchsbaum2000ColorICA,Sejnowski2001SparseCoding_Chromatic,Sejnowski2002SparseCoding_Chromatic,Sejnowski2003ColorImagesICA,Caywood2004SparseCoding_Chromatic}. One common observation in these studies is that the learned visual features (eigenvectors or independent components) segregate into black/white, blue/yellow, and red/green opponent structures, either shaped as Fourier basis functions or Gabor kernel functions~\cite{Sejnowski2001SparseCoding_Chromatic}.  These results are similar to what is obtained with ZCA (Zero-component analysis)~\cite{Sejnowski2003ColorImagesICA}, although ZCA has small connections to all of the inputs, since it does not inherently try to minimize connections.

\paragraph*{Grayscale Videos} \label{sec:spca:experiments:video}
To explore the spatio-temporal structure of natural videos, we collected a video dataset of $27$ clips from nature documentaries. Just as a two dimensional image patch can be flattened into a vector of input responses, a three dimensional spatiotemporal patch of video can also be tranformed into a vector. These vectors can then be given to the model as input. 

\begin{figure}
    \begin{center}
     \subfigure[two video filters learned by Sparse PCA (time from left to right)]{\includegraphics[width=0.9\textwidth]{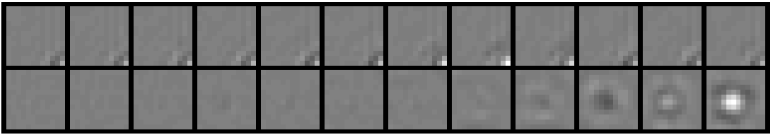}\label{fig:spca:video}} \\
\subfigure[distribution of video features learned by Sparse PCA]{\includegraphics[width=0.8\textwidth]{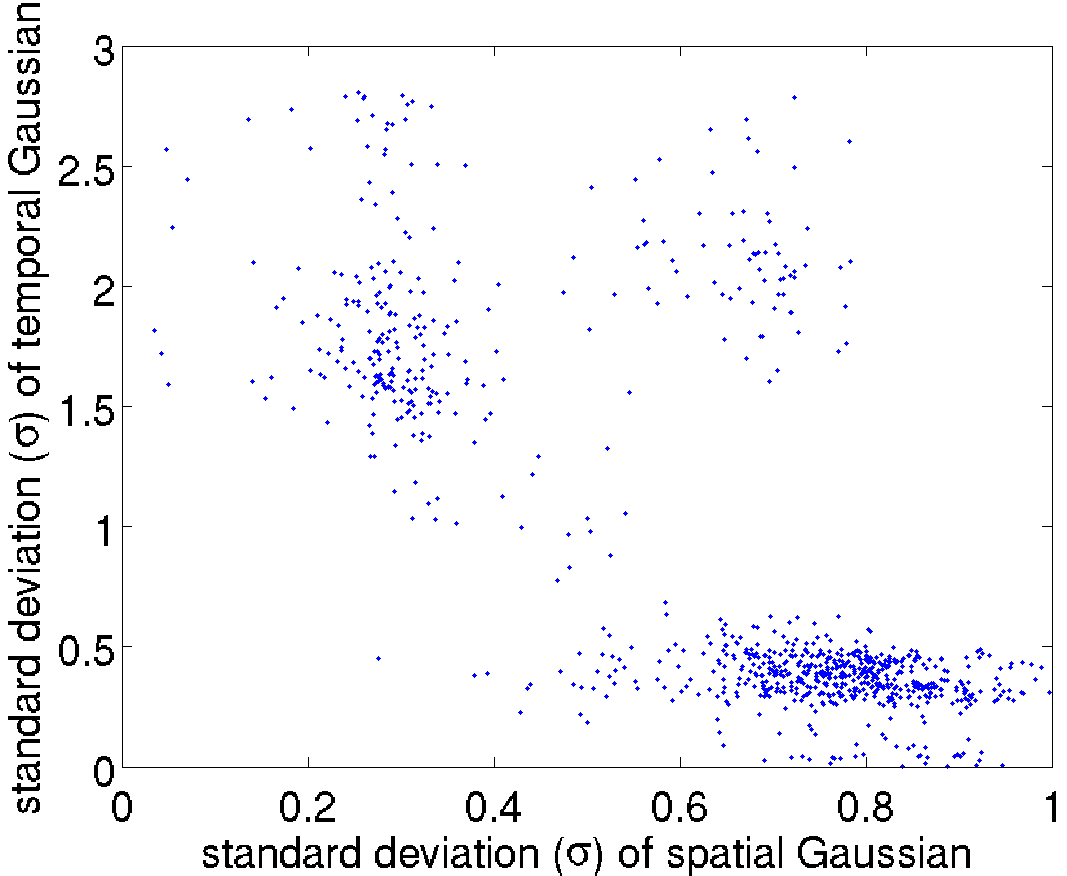}\label{fig:spca:video:dist}} \\
       \end{center}
    \caption{Video features segregate into two groups: small, persistent features, and large, brief features.}    
\end{figure}

The learned features are black/white blobs whose contrast changes over time. As shown in Figure~\ref{fig:spca:video}, the features can be well fitted to spatio-temporal Gaussians. The corresponding filters are spatially center-surround shaped. Their temporal profile seems to provide an ``edge'' detector along the temporal axis, which is similar to the temporal profile describing retinal ganglion cells~\cite{Chichilnisky2008PredictRGC}. To see the animation file of the learned filters, see \url{http://cseweb.ucsd.edu/~gary/video_W.gif}.

Another interesting observation is that most of  the features segregate into two groups: those with low-spatial and high-temporal frequencies (centered around (0.3,1.75) in the figure), and those with high-spatial and low-temporal frequencies (centered around (.75,0.4) in the figure), as plotted in Figure~\ref{fig:spca:video:dist}. This suggests that the division of ganglion cells into the magno-pathway and the parvo-pathway represents an efficient encoding of the visual environment.
This segregation appears to reveal statistical properties of natural videos, instead of coming from our specific algorithm. In fact, we found that this segregation exists even for features learned with standard PCA. As in PCA of static images, however, the features are not biologically plausible. 

\paragraph*{Sound} \label{sec:spca:experiments:sound}

\begin{figure}
    \begin{center}
    \subfigure[revcor filter from cat]{\includegraphics[width=0.9\textwidth]{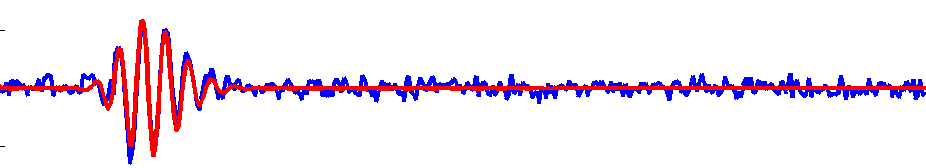}\label{fig:spca:auditory:cat2}}\\
    \subfigure[revcor filter from cat]{\includegraphics[width=0.9\textwidth]{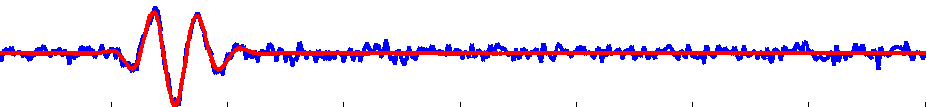}\label{fig:spca:auditory:cat3}}\\
    \subfigure[filter learned from the TIMIT dataset]{\includegraphics[width=0.9\textwidth]{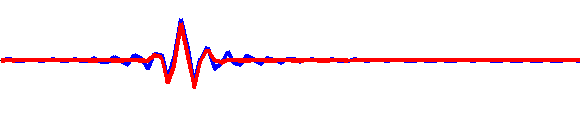}\label{fig:spca:auditory:timit}}\\    
    \subfigure[filter learned from the Pittsburgh dataset]{\includegraphics[width=0.9\textwidth]{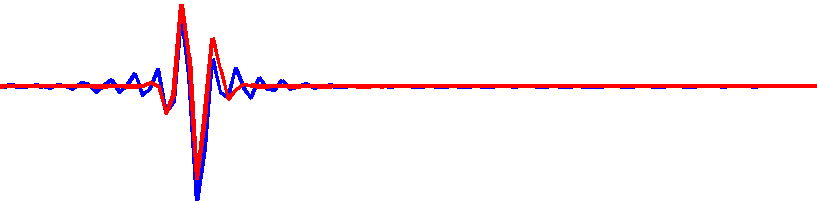}\label{fig:spca:auditory:pitts}}\\
\end{center}
    \caption{Figure~\ref{fig:spca:auditory:cat2} and \ref{fig:spca:auditory:cat3} plot the revcor filters estimated from cat's auditory nerve fibers using the linear reverse correlation technique, as well as the fitted gammatone filter.  Figure~\ref{fig:spca:auditory:timit} and \ref{fig:spca:auditory:pitts} plot filters learned from the TIMIT speech dataset and the Pittsburgh environmental sound dataset respectively. The blue line plots the estimated filter; the red line plots the fitted gammatone filter.}
    \label{fig:spca:auditory}
\end{figure}

We applied our algorithm to three sound datasets and get qualitatively similar results on these three datasets. As shown in Figure~\ref{fig:spca:auditory}, the learned filters can be well fitted to gammatone filters. Gammatone filters resemble the filtering properties of auditory nerve fibers estimated using the reverse correlation technique from animal such as cats~\cite{Carney1990Cochlear} and chinchillas~\cite{Ruggero2005Cochlear}. 

Note that this is (to the best of our knowledge) the first time a non-ICA algorithm has learned gammatone-like filters from sound. Also, since we used the same algorithm for both visual and auditory modalities, this provides an answer to the question posed by Olshausen and O'Connor (2002): ``Perhaps an even deeper question is why ICA accounts for neural response properties at the very earliest stage of analysis in the auditory system, whereas in the visual system ICA accounts for the response properties of cortical neurons, which are many synapses removed from photoreceptors.''  Our model suggests that it is not necessary to use ICA to obtain gammatone filters from sound; rather, Sparse PCA can account for the receptive fields of neurons at the very earliest stages of analysis in both auditory and visual modalities.

\section*{Discussion} \label{sec:spca:discussion}

We have suggested three principles that can be used to explain precortical encoding: information preservation, minimization of the neural wiring, and response equalization. Each of these principles can be independently justified via evolutionary and energy minimization arguments. Clearly, an organism should try to extract as much relevant information as possible from its environment. As organisms evolve to survive in more enriched environments, which information is relevant becomes more difficult to encode in the genome. As Barlow and Attneave have argued, redundancy reduction for efficient coding is a reasonable response to environmental complexity. Minimizing energy usage suggests constructing the minimal architecture possible, via minimizing wiring, which is a win in terms of both development and daily energy budgets. Finally, equalizing the work (response normalization) results in no single component being crucial to the organism.  

The model is closely related to previous theories, but differs in crucial respects. We were inspired by Vincent et al.'s model \cite{Vincent2005Retina}, which also attempted to derive center surround receptive fields by information preservation and minimizing wiring. By including the response normalization constraint, we were able to obtain a more stable algorithm with none of the occasional double receptive fields generated by their model. Unlike many previous models, explicitly decorrelating the outputs of our model was not necessary in order to obtain our results. Hence the model integrates two of the three components suggested as being necessary for any retinal coding model by Graham,  Chandler, and Field \cite{Field2006WhiteningTheory}  - response equalization and sparseness, but replaces decorrelation with minimization of the neural wiring.  The Sparse PCA model has an interesting link to Olshausen's sparse coding model; both models try to minimize the reconstruction error with response equalization, but our model imposes sparseness on the dictionary, while the sparse coding model imposes sparseness on the output. Finally, the model is closely related to that described in \cite{Doi2007RetinalCoding}, where they proposed a model in which the retinal output is a linear transform of the input $\mathbf{s}=\mathbf{W}\mathbf{x}$, ignoring optical blur. The objective of their model is to minimize the difference between the input $\mathbf{x}$ and its reconstruction $\mathbf{A}\mathbf{W}\mathbf{x}$. Hence their model can be seen as the approximation we take when $\lambda$ is small and $L > M$ (described in the Supplementary materials). They also include two extra terms during learning, to regularize the average amplitude of the outputs, and to impose sparseness on $\mathbf{W}$ (instead of $\mathbf{A}$). Our version, with sparseness on $\mathbf{A}$, leads to a simple interpretation of locally-connected ganglion cells with an inhibitory surround. Also, the convexity properties of our model lead to excellent convergence properties. 

We derived an efficient algorithm to learn the model parameters by transforming it into a sparse coding problem. Our approximate algorithm uses only the covariance matrix C, and runs orders of magnitude faster than the exact algorithm, while obtaining results that are less than one percent different in the objective function and learned basis functions. Our approximation works well under the assumption that $\lambda$ is small and $L > M$. While fast approximation is not a necessary part of a successful model, the speed of computation allows it to be used on more rich data, such as video.

We applied our algorithm to grayscale
images, color images, grayscale videos, human speech, and environmental
sound, and learned visual and auditory filters that resemble the
filtering properties of retinal ganglion cells and auditory nerve
fibers. Some of the learned filters are novel. For example, it learns
the magno and parvo segregation pathways; and it learns the
gammatone filters from natural sound. 

Finally, as noted above, our model suggests an answer the question of why ICA gives features corresponding to cortical receptive fields in vision, but seems necessary to obtain gammatone-like filters for sound, which is a pre-cortical level of processing. Our suggestion is that ICA is not necessary to obtain gammatone filters; rather, a PCA algorithm with sparsity and response equalization constraints can result in gammatone filters for sound, while also producing receptive fields similar to peripheral neurons in vision.

\section*{Materials and Methods}
\paragraph*{Model}
As previously discussed, our model seeks to find the output vectors $\mathbf{s} \in R^M$ (the signal from the retinal ganglion cells) and basis functions $A \in R^{L \times M}$  such that the following objective function is minimized:
\begin{eqnarray}
E = \left< \frac{\|\mathbf{x} - \mathbf{A}\mathbf{s}\|_2^2}{2} \right> + \lambda \|\mathbf{A}\|_1 \label{eq:spca:original_objective}
\end{eqnarray}
(where $\left< \cdot \right>$ denotes taking average over all the input samples) subject to the constraint that  the average output of each cell  $\left<s_i^2\right> \leq 1$. Upon convergence, the model will satisfy  $\left<s_i^2\right> = 1$ as otherwise the objective function could easily be further reduced.

Vincent et al.\ interpret $\mathbf{A}$ as the synaptic strength between the input and output neurons; and hence they interpret the sparsity penalty on $\mathbf{A}$ as the desire to minimize the neural wiring cost. From a generative point of view, the columns of $\mathbf{A}$ are the elementary features that the model uses to ``construct'' the observed inputs. Thus the overall objective function can be understood as capturing most information of the inputs using an economic dictionary of elementary features. However, when $\mathbf{A}$ is fixed and full rank, the optimal output is in fact given by $\mathbf{s}^* = \mathbf{Wx}$, where $\mathbf{W}$ is the pseudoinverse of $\mathbf{A}$ (if we ignore our constraint on the average activation of a cell - an assumption we experimentally verified as yielding a suitable approximate solution). If we apply the linear reverse correlation technique to the model neurons~\cite{Chichilnisky2001ReverseCorrelation}, which recovers the filters transforming the inputs to the outputs, we will get the rows of $\mathbf{W}$  as the model neurons' receptive fields.  

The columns of $\mathbf{A}$ and the rows of $\mathbf{W}$ describe different aspects of the model. The columns of $\mathbf{A}$ describe the elementary \emph{features} that the model uses to construct the observed inputs; this forms a kind of visual dictionary. The rows of $\mathbf{W}$, on the other hand, represent the best linear \emph{filters} to recover the weights assigned to the elementary features when generating the observed inputs. Since $\mathbf{W}$ better reflects the properties of cells' receptive fields, we focus primarily on $\mathbf{W}$ throughout the paper.

Our objective function allows us to derive an efficient algorithm to learn the model parameters. Our objective function can be written as: \begin{eqnarray}
E = \frac{\| \mathbf{X} - \mathbf{A} \mathbf{S}\|_F^2}{2 n} + \lambda \|\mathbf{A}\|_1 \label{eq:spca:objective:matrix}
\end{eqnarray}
where the columns of $\mathbf{X}$ and $\mathbf{S}$ store the input and output vectors; $\|\|_F$ denotes taking the Frobenius norm of a matrix (i.e., the square root of the sum of squared entries); $n$ denotes the number of samples. Our constraint now is that each row of $\mathbf{S}$ is constrained to have L2 norm less than or equal to $\sqrt{n}$.

As shown by \cite{Bach2010MatrixFactorization}, the Sparse Coding problem~\cite{Olshausen1996SparseCoding_SimpleCell} can be expressed in the matrix factorization form as:
\begin{eqnarray}
E = \frac{\| \mathbf{X} - \mathbf{A} \mathbf{S}\|_F^2}{2} + \lambda \|\mathbf{S}\|_1 \label{eq:spca:sc_objective}
\end{eqnarray}
Each column of $\mathbf{A}$ is constrained to have L2 norm less than or equal to $q$. Sparse PCA can then utilize Sparse Coding algorithms because the objective function in Eq~(\ref{eq:spca:objective:matrix}) can be re-written as:
\begin{eqnarray}
E = \frac{\| \mathbf{X}^t - \mathbf{S}^t \mathbf{A}^t\|_F^2}{2 n} + \lambda \|\mathbf{A}^t\|_1 \label{eq:spca:objective:matrix2}
\end{eqnarray}
Now we see that Eq~(\ref{eq:spca:objective:matrix2}) can be symbolically mapped to Eq~(\ref{eq:spca:sc_objective}), if we replace $\mathbf{X}^t$ with $\mathbf{X}$, $\mathbf{S}^t$ with $\mathbf{A}$, $\mathbf{A}^t$ with $\mathbf{S}$, with extra care taken to deal with $n$ and $\sqrt{n}$.

The above discussion suggests another interpretation of the retinal coding model: it can be interpreted as removing redundancy between input samples, because Sparse Coding is usually interpreted as removing redundancy between input dimensions~\cite{Lewicki1999ICA_Bayesian,Lewicki2000ICA_Overcomplete}. In this sense, Architecture-1 ICA~\cite{Bartlett1998ICA_Face}, which applies ICA to $\mathbf{X}^t$ instead of $\mathbf{X}$, was perhaps the first Sparse PCA algorithm ever proposed.

We then only need to pick some efficient Sparse Coding algorithms to optimize our model parameters. Typically, optimizing the objective function in Eq~(\ref{eq:spca:sc_objective}) is factored into two sub-problems: optimizing $\mathbf{A}$ while fixing $\mathbf{S}$, and optimizing $\mathbf{S}$ while fixing $\mathbf{A}$. Both sub-problems are convex optimization problems (this is one reason we impose $\left<s_i^2\right> \leq 1$ instead of $\left<s_i^2\right> = 1$; otherwise optimizing $\mathbf{S}$ is no longer a convex optimization problem). Recently, it was shown that the coordinate descent algorithm is considerably faster than competing methods for both sub-problems~\cite{Friedman2010CoordinateDescent,Bach2010MatrixFactorization}. These algorithms are described in Appendix~\ref{sec:spca:appendix:coordinate_descent}.

The above algorithm has a computational complexity that depends on the number of samples $n$. Here we give an approximate algorithm whose complexity only relies on $L$ and $M$, the input and output dimensionalities. In our experiments on grayscale images, this reduces the computation time from $37$ minutes to $10$ seconds, with the learned parameters very close to those learned without approximation. The derivation of the algorithm also helps us to see the connection between this retinal coding model and the output decorrelation theory~\cite{Atick1992RetinaDecorrelation}.

Our algorithm utilizes two approximations, both of which rely on the condition that $\lambda$ is small and $L > M$. Under such a condition, the first approximation we use is to calculate the optimal output $\mathbf{s}$ using (See Appendix~\ref{sec:spca:appendix:approximate} for detailed derivations):
\begin{eqnarray}
\mathbf{s}^* \approx \mathbf{W} \mathbf{x} \quad \mbox{($\mathbf{W} = (\mathbf{A}^t \mathbf{A})^{-1} \mathbf{A}^t$)}
\end{eqnarray}
Replacing the above approximation into the original objective function, we get
\begin{eqnarray}
E \approx \frac{\Tr \left( (\mathbf{I} - \mathbf{A} \mathbf{W}) \mathbf{C} (\mathbf{I} - \mathbf{A} \mathbf{W})^t \right)}{2} + \lambda \|\mathbf{A}\|_1 \label{eq:spca:objective:approx1}
\end{eqnarray}
where $\mathbf{C} = \left<\mathbf{x}\mathbf{x}^t \right>$, $\mathbf{I}$ denotes the identity matrix, and $\Tr$ denotes taking the trace of a matrix. The constraint $\left< s_i^2 \right> \leq 1$ can be expressed as $\diag \left( \mathbf{W} \mathbf{C} \mathbf{W}^t \right) \leq \mathbf{1}$.

Since $\mathbf{C} = \left<\mathbf{x}\mathbf{x}^t\right>$ is positive semidefinite, we can factor it using the eigenvalue decomposition $\mathbf{C} = \mathbf{U} \mathbf{V} \mathbf{U}^t$, where $\mathbf{U}$ is a unitary matrix (i.e., $\mathbf{U} \mathbf{U}^t = \mathbf{I}$) containing the eigenvectors as its columns; $\mathbf{V}$ is a diagonal matrix with the eigenvalues on its diagonal. Let $\mathbf{B} = \mathbf{U} \mathbf{V}^{1/2}$, we get $\mathbf{C} = \mathbf{B}\mathbf{B}^t$. Replacing $\mathbf{C} = \mathbf{B}\mathbf{B}^t$ into Eq~(\ref{eq:spca:objective:approx1}), we get
\begin{eqnarray}
E &\approx& \frac{\| \mathbf{B}^t - \mathbf{Z}^t \mathbf{A}^t\|_F^2}{2} + \lambda \|\mathbf{A}^t\|_1 \label{eq:spca:final_objective}
\end{eqnarray}
where $\mathbf{Z} = \mathbf{W} \mathbf{B}$. The second approximation we use is to relax $\mathbf{Z}$ to a free variable instead of constraining it to $\mathbf{Z} = \mathbf{W} \mathbf{B}$. The constraint becomes that each column of $\mathbf{Z}^t$ should have L2 norm less than or equal to $1$.

Now we see that Eq~(\ref{eq:spca:final_objective}) can also be symbolically mapped to Eq~(\ref{eq:spca:sc_objective}), if we replace $\mathbf{B}^t$ with $\mathbf{X}$, $\mathbf{Z}^t$ with $\mathbf{A}$, and $\mathbf{A}^t$ with $\mathbf{S}$. Hence the objective function in Eq~(\ref{eq:spca:final_objective}) can also be optimized by efficient Sparse Coding algorithms, such as the coordinate descent algorithms in Appendix~\ref{sec:spca:appendix:coordinate_descent}. Since Sparse Coding reduces the redundancy between inputs, the above derivation also brings up another interpretation of the retinal coding model: it can be seen as removing the redundancy between the eigenvectors of $\mathbf{C}$ when $\lambda$ is small and $L > M$.

Our algorithm efficiently minimizes the objective function. We test two optimization methods: directly optimize the objective function in Eq~(\ref{eq:spca:objective:matrix2}), or first initialize $\mathbf{A}$ by optimizing Eq~(\ref{eq:spca:final_objective}) and then optimize Eq~(\ref{eq:spca:objective:matrix2}). We implement the experiments in single precision on a computer server with Intel Core i7 processors. When we reduce the dimensionality from $400$ to $100$ and set $\lambda=0.004$, it takes $37$ minutes to directly optimize Eq~(\ref{eq:spca:objective:matrix2}). On the other hand, it takes less than $10$ seconds to initialize $\mathbf{A}$ by optimizing Eq~(\ref{eq:spca:final_objective}) and another $40$ seconds to optimize Eq~(\ref{eq:spca:objective:matrix2}). As shown in Figure~\ref{fig:spca:grayscale:method:objective}, our approximate method efficiently minimizes the objective function. In fact, after $\mathbf{A}$ is initialized with the approximate method, further optimizing the objective function with the direct method only causes a less than $0.1\%$ change of the objective function, on average the weights in $\mathbf{A}$ are only changed by $0.4\%$, and no visible difference in the features and filters can be observed. The closeness of the approximation means that in later experiments (color images, video, and sound) the approximation provided by Eq~(\ref{eq:spca:final_objective}) is used without additional optimization. 

\paragraph*{Grayscale Images}
The four image sets used were van Hateren natural images~\cite{Hateren1998SimpleCellICA}, Kyoto natural images~\cite{Sejnowski2003ColorImagesICA}, Berkeley segmentation dataset~\cite{BSDS300}, and Caltech-256 object category dataset~\cite{Caltech256}. We didn't observe any qualitative difference between the features learned from these datasets. The features reported are obtained using van Hateren natural images. We use a subset of it selected by \cite{Lewicki2009ComplexCells}, which contains 110 $1536 \times 1024$ grayscale images.

For each image, we discard two pixels off the image borders, normalize the pixel values to $[0,1]$, then apply a nonlinear function that simulates the cone processing~\cite{Lewicki2009ComplexCells}:
\begin{eqnarray}
x = 1 - \exp(-k \cdot x) \label{eq:spca:cone_nonlinearity}
\end{eqnarray}
where $x$ denotes the pixel value, and $k$ is selected for each image such that its average pixel value equals $0.5$ after the nonlinearity. This nonlinearity does not seem to alter the features learned by the retinal coding model, but might help to expose higher-order statistical structure~\cite{Lewicki2009ComplexCells}. Then we randomly sample $1000$ $20 \times 20$ image patches from each image. Figure~\ref{fig:spca:grayscale:method} compares the distributions of connection weights between PCA and SPCA. 

\paragraph*{Chromatic Images}
We applied our algorithm to four chromatic image datasets: the Kyoto natural image dataset~\cite{Sejnowski2003ColorImagesICA}, the McGill color image dataset~\cite{McGillColorImages}, the Berkeley segmentation dataset~\cite{BSDS300}, and the Caltech-256 object category dataset~\cite{Caltech256}. From these datasets we learn qualitatively similar features. Below we report the features learned from Kyoto image dataset, which contains 62 $1000 \times 1280$ pixel chromatic images of natural scenes.

First, we estimate the retinal L, M, S cones' responses to those images. The original images are stored in sRGB color representation~\cite{sRGB1996}. We normalize the pixel values of each image to $[0,1]$, then transform the image to the CIE XYZ color space~\cite{XYZ1931}, from which we estimate the LMS cone responses following the CIECAM02 color appearance model~\cite{CIECAM02}. In this manner, we can roughly estimate how the retinal L, M, and S cones would respond when presented with the image content.

After that, we apply the cone nonlinearity in Eq~(\ref{eq:spca:cone_nonlinearity}) to the estimated LMS cone responses. Then we extract all $20$ by $20$ image patches, and estimate the matrix $\mathbf{C}$. We apply the retinal coding model using the approximation in Eq~(\ref{eq:spca:final_objective}) to reduce the dimensionality from $1200$ to $256$ with $\lambda=0.002$. The resulting model captures $99.75\%$ of the variance captured by an optimal linear model (i.e., standard PCA) with $256$ output neurons, with $96.11\%$ of its connections being absolute zero.

\paragraph*{Grayscale Videos}
To explore the spatio-temporal structure of natural videos, we collected a video dataset of $27$ clips from YouTube (see online supporting material for their URLs). All the videos are from natural history shows from the BBC World Wide channel (\url{www.youtube.com/user/BBCWorldwide}). In order to have a realistic sample of natural videos, we eliminate those that are obviously unnatural, such as walking dinosaurs, cavemen, or pigeon-mounted cameras. We calculated the power spectrum of the videos in order to eliminate interlaced-format videos, which have an ellipse-shaped power spectrum elongated in the horizontal direction. We initially applied our algorithm to van Hateren video dataset~\cite{Hateren1998TemporalICA}. The learned features and filters are qualitatively similar to those reported below, except that some of the learned features are ellipse or even bar shaped and are elongated along the horizontal direction. These features most likely result from the fact that the original videos are interlaced (i.e., recording only the odd numbered lines in one frame, and the even numbered lines in the next frame). Although the videos were de-interlaced by block averaging with $2 \times 2 \times 2$ (van Hateren, personal communication), their spatial power spectrum is still ellipse-shaped with more energy along the horizontal direction.

The original YouTube videos are in color. We transform them to grayscale videos using the method described by the Matlab function rgb2gray, normalize the pixel values between $[0,1]$, and apply the cone nonlinearity in Eq~\ref{eq:spca:cone_nonlinearity}. Then we estimate the correlation matrix $\mathbf{C}=\mathbf{x}\mathbf{x}^t$ for all the $12 \times 12 \times 12$ video cubes (with the local mean removed - keeping in the local mean did not substantially change the qualitative results, but it did seem to produce slightly noisier filters), and apply our algorithm to $\mathbf{C}$ using the approximate method.

\paragraph*{Sound}
We applied our algorithm to three sound datasets: the Pittsburgh natural sounds dataset~\cite{Lewicki2006EfficientAuditoryCoding}, the TIMIT speech dataset~\cite{Lamel1986TIMIT}, and rainforest mammal vocalization dataset~\cite{RainforestMammal}. The Pittsburgh natural sounds dataset contains $48$ recordings of natural sound recorded around the Pittsburgh region, including ambient sounds  (such as rain, wind, and streams) and quick acoustic events (such as snapping twigs, breaking wood, and rock impacts). The TIMIT speech dataset contains English speech from $630$ speakers, with each person speaking $10$ sentences. The rainforest mammal vocalization dataset contains the characteristic sounds of $109$ species of rainforest mammals, such as primates, anteaters, bats, jaguars, and manatees.

For each dataset, each recording is re-sampled at $16$ kHz. We normalize the maximum amplitude of each recording to $1$, take all the segments formed using a sliding window of $128$ sample points (about a $8$ millisecond window), then estimate the matrix $\mathbf{C}$ for the sound segments. We then apply the approximate method to learn the features and filters.

\bibliography{./reference}

\section*{Acknowledgements}

We thank 
Lingyun Zhang,
Wensong Xu, 
and Chris Kanan 
for helpful discussions, 
Ben Vincent for sharing the source code of his model, 
Yan Karklin for sharing his image preprocessing code, 
Eizaburo Doi for sharing his code to calculate LMS cone representations, 
Vivienne Ming 
and Mike Lewicki 
for providing the auditory data, 
Malcolm Slaney for suggesting the experiments on noise images, and Terry Sejnowski for sharing computational resources.  This work was supported in part by NSF Science of Learning Center grants SBE-0542013 and SMA-1041755 to the Temporal Dynamics of Learning Center, NSF grant IIS-1219252 to GWC. 

\listoffigures

\newpage
\centerline{\large Supporting Online Material for}

\centerline{A single model explains both visual and auditory precortical coding}
\centerline{Honghao Shan, Matthew H. Tong, Garrison W. Cottrell}

\appendix
\section{Coordinate Descent Algorithms} \label{sec:spca:appendix:coordinate_descent}

Suppose we want to optimize an objective function $f(\mathbf{x})$, where $\mathbf{x}$ is a high dimensional vector. To do so, the coordinate descent algorithm cyclically optimizes each dimension of $\mathbf{x}$. Each time, it optimizes one dimension of $\mathbf{x}$ while fixing the other dimensions. Once this dimension is optimized, the algorithm optimizes another dimension. This process repeats until the objective function can no longer be optimized.

Here we list the coordinate descent algorithm for optimizing the Sparse Coding objective function:
\begin{eqnarray}
E = \left< \frac{1}{2} \|\mathbf{x}-\mathbf{A}\mathbf{s}\|_2^2 + \lambda \|\mathbf{s}\|_1 \right>
\end{eqnarray}
Each column of $\mathbf{A}$ is constrained to have L2 norm less than or equal to $q$.

\paragraph{Optimizing $\mathbf{s}$ While Fixing $\mathbf{A}$} When we try to optimize the $i$-th coordinate of $\mathbf{s}$, with $\mathbf{A}$ and all the other coordinates of $\mathbf{s}$ being fixed, the optimal $s_i$ is given by~\cite{Friedman2010CoordinateDescent}:
\begin{eqnarray}
y &=& \mathbf{a}_i^t \mathbf{x} - \sum_{j \neq i} \mathbf{a}_i^t \mathbf{a}_j s_j \\
s_i^* &=& \left\{ \begin{array}{cl}
  (y - \lambda) / \|\mathbf{a}_i\|_2^2, & \mbox{if $y > \lambda$;} \\
  (y + \lambda) / \|\mathbf{a}_i\|_2^2, & \mbox{if $y < - \lambda$;} \\
  0, & \mbox{otherwise.}
\end{array} \right.
\end{eqnarray}
where $\mathbf{a}_i$ denotes the $i$-th column of $\mathbf{A}$.

\paragraph{Optimizing $\mathbf{A}$ While Fixing $\mathbf{s}$} When we try to optimize the $i$-th column of $\mathbf{A}$, with $\mathbf{s}$ and all the other columns of $\mathbf{A}$ being fixed, the optimal $\mathbf{a}_i$ is given by~\cite{Bach2010MatrixFactorization}:
\begin{eqnarray}
\mathbf{u} &=& \left<s_i \mathbf{x} \right> - \sum_{j\neq i} \mathbf{a}_j \left<s_j s_i\right> \\
\mathbf{a}_i^* &=& \frac{\mathbf{u}}{\max(\left<s_i^2\right>, \|\mathbf{u}\|_2/q)}
\end{eqnarray}

\begin{figure}
    \begin{center}
    \subfigure[$\mathbf{W}\mathbf{x}$ versus $\mathbf{s}$]{\includegraphics[width=0.34\textwidth]{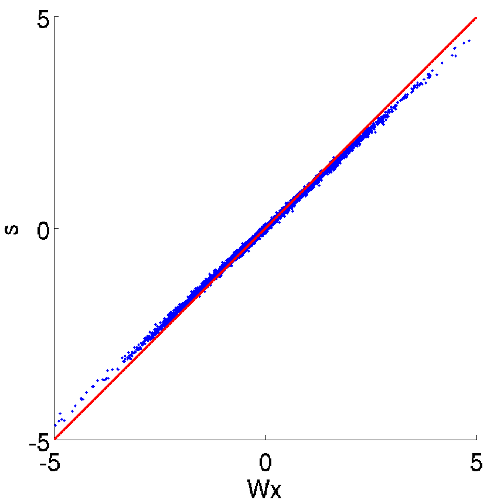}\label{fig:spca:grayscale:reverse_correlation:Wx}}
    \subfigure[Linear reverse correlation]{\includegraphics[width=0.3\textwidth]{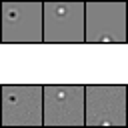}\label{fig:spca:grayscale:reverse_correlation:XS}}
    \end{center}
    \caption{$\mathbf{W}$ approximates the model neurons' receptive fields. In Figure~\ref{fig:spca:grayscale:reverse_correlation:Wx}, we compare $\mathbf{W} \mathbf{x}$ with the optimal $\mathbf{s}$ inferred using the direct method during the experiments on grayscale image patches. If $\mathbf{s} = \mathbf{W} \mathbf{x}$ holds exactly, all the points should lie on the red solid line. Once we have learned the optimal $\mathbf{A}$ on grayscale image patches, we feed the model with white noise inputs and infer their optimal outputs, then use $\left<\mathbf{x}\mathbf{s}^t\right>$ to estimate the model neurons' receptive fields. The top panel in Figure~\ref{fig:spca:grayscale:reverse_correlation:XS} displays three rows of $\mathbf{W}$; the bottom panel displays the estimated receptive fields for the corresponding neurons.}
    \label{fig:spca:grayscale:reverse_correlation}
\end{figure}

\section{Derivation of the approximate algorithm} \label{sec:spca:appendix:approximate}

The revised retinal coding model aims to minimize the following objective function:
\begin{eqnarray}
E = \left< \frac{\|\mathbf{x} - \mathbf{A}\mathbf{s}\|_2^2}{2} \right> + \lambda \|\mathbf{A}\|_1
\end{eqnarray}
subject to the constraint that
\begin{eqnarray}
\left< s_i^2 \right> \leq 1 \quad \mbox{for every $i$}
\end{eqnarray}

Below we show how to approximate the above objective function when $\lambda$ is small and $L > M$ (i.e., when we are reducing the data dimensionality).

\paragraph{The First Approximation} Let's first check the functional roles of the three terms in the objective function: the reconstruction error, the sparsity penalty, and the constraint.

If the objective function only contains the reconstruction error term, the model is reduced to the standard PCA problem~\cite{Jolliffe2002BookPCA}. The optimal basis functions (i.e., the columns of $\mathbf{A}$) should span the subspace spanned by the eigenvectors of $\left<\mathbf{x}\mathbf{x}^t \right>$ with top eigenvalues. The optimal outputs are given by $\mathbf{s}^* = \mathbf{W} \mathbf{x}$, where $\mathbf{W}$ is the pseudo-inverse of $\mathbf{A}$. Changing the basis functions' individual directions and lengths within the subspace won't change the reconstruction error, because for any full-rank matrix $\mathbf{G} \in R^{M \times M}$ we have
\begin{eqnarray}
\mathbf{A} \mathbf{s} = (\mathbf{A} \mathbf{G}) (\mathbf{G}^{-1} \mathbf{s})
\end{eqnarray}
That is, for any $\mathbf{A}_{new} = \mathbf{A} \mathbf{G}$ which spans the same subspace as $\mathbf{A}$ but with different individual lengths and directions, we can find $\mathbf{s}_{new} = \mathbf{G}^{-1} \mathbf{s}$ such that the reconstruction error remains to be the minimum.

The objective function is not changed by adding the constraint $\left< s_i^2 \right> \leq 1$. For any value of $\mathbf{A}$ and $\mathbf{s}$, we can always divide $s_i$ (the $i$-th coordinate of $\mathbf{s}$) and multiply $\mathbf{a}_i$ (the $i$-th column of $\mathbf{A}$) with some value $\alpha$ to satisfy the constraint without changing the reconstruction error. In other words, the constraint term only specifies the lengths of the basis functions.

The sparsity penalty $\|\mathbf{A}\|_1$ will shrink the basis functions' lengths and rotate their directions. However, when $\lambda > 0$ is sufficiently small such that the subspace that the basis functions span is  mainly determined by the reconstruction error, the sparsity penalty will only serve to rotate the basis functions within the subspace determined by the reconstruction error.

Our first approximation utilizes the above analysis. When $\lambda$ is small and $L > M$, we use $\mathbf{s}^* \approx \mathbf{W} \mathbf{x}$, where $\mathbf{W}$ is the pseudo-inverse of $\mathbf{A}$. Substituting it into the objective function in Eq~(\ref{eq:spca:original_objective}), we get:
\begin{eqnarray}
E &\approx& \left< \frac{\|\mathbf{x} - \mathbf{A} (\mathbf{W} \mathbf{x})\|_2^2}{2} \right> + \lambda \|\mathbf{A}\|_1 \\
&=& \left< \frac{\Tr \left((\mathbf{x} - \mathbf{A} \mathbf{W} \mathbf{x}) (\mathbf{x} - \mathbf{A} \mathbf{W} \mathbf{x})^t \right) }{2} \right> + \lambda \|\mathbf{A}\|_1 \label{eq:spca:trace} \\
&=& \frac{\Tr \left( (\mathbf{I} - \mathbf{A} \mathbf{W}) \left<\mathbf{x}\mathbf{x}^t \right> (\mathbf{I} - \mathbf{A} \mathbf{W})^t \right)}{2} + \lambda \|\mathbf{A}\|_1 \\
&=& \frac{\Tr \left( (\mathbf{I} - \mathbf{A} \mathbf{W}) \mathbf{C} (\mathbf{I} - \mathbf{A} \mathbf{W})^t \right)}{2} + \lambda \|\mathbf{A}\|_1 \quad \mbox{(let $\mathbf{C} = \left<\mathbf{x}\mathbf{x}^t \right>$)} \label{eq:spca:appendix:objective:approx1}
\end{eqnarray}
where $\mathbf{I}$ denotes the identity matrix, $\Tr$ denotes taking the trace of a matrix. The constraint can now be approximated as:
\begin{eqnarray}
\left< s_i^2 \right> \approx \left< (\mathbf{w}_i \mathbf{x}) (\mathbf{w}_i \mathbf{x})^t \right> = \mathbf{w}_i \left< \mathbf{x} \mathbf{x}^t \right> \mathbf{w}_i^t \leq 1, \quad \mbox{or} \quad \diag(\mathbf{W} \mathbf{C} \mathbf{W}^t) \leq \mathbf{1} \label{eq:spca:si2WC}
\end{eqnarray}
where $\mathbf{w}_i$ denotes the $i$-th row of $\mathbf{W}$, and $\mathbf{1}$ denotes a vector of $1$'s. Hence, when $\lambda$ is small and $L > M$, the model mainly serves to capture the second-order statistical structure of the inputs.

\paragraph{The Second Approximation} Since $\mathbf{C} = \left<\mathbf{x}\mathbf{x}^t\right>$ is positive semidefinite, we can factor it using the eigenvalue decomposition $\mathbf{C} = \mathbf{U} \mathbf{V} \mathbf{U}^t$, where $\mathbf{U}$ is a unitary matrix (i.e., $\mathbf{U} \mathbf{U}^t = \mathbf{I}$) containing the eigenvectors as its columns; $\mathbf{V}$ is a diagonal matrix with the eigenvalues on its diagonal. Let $\mathbf{B} = \mathbf{U} \mathbf{V}^{1/2}$, we get $\mathbf{C} = \mathbf{B}\mathbf{B}^t$. Substituting this into the objective function in Eq~(\ref{eq:spca:appendix:objective:approx1}), yields
\begin{eqnarray}
E &=& \frac{\Tr \left( (\mathbf{I} - \mathbf{A} \mathbf{W})  \mathbf{B}\mathbf{B}^t (\mathbf{I} - \mathbf{A} \mathbf{W})^t \right)}{2} + \lambda \|\mathbf{A}\|_1  \\
&=& \frac{\Tr \left( (\mathbf{B} - \mathbf{A} \mathbf{W} \mathbf{B}) (\mathbf{B} - \mathbf{A} \mathbf{W}\mathbf{B})^t \right)}{2} + \lambda \|\mathbf{A}\|_1 \\
&=& \frac{\| \mathbf{B} - \mathbf{A} \mathbf{W} \mathbf{B}\|_F^2}{2} + \lambda \|\mathbf{A}\|_1 \\
&=& \frac{\| \mathbf{B} - \mathbf{A} \mathbf{Z} \|_F^2}{2} + \lambda \|\mathbf{A}\|_1 \quad \mbox{(let $\mathbf{Z} = \mathbf{W} \mathbf{B}$)}
\end{eqnarray}
The constraint becomes that each row of $\mathbf{Z}$ should have L2 norm less than or equal to $1$:
\begin{eqnarray}
\diag(\mathbf{W} \mathbf{C} \mathbf{W}^t) = \diag(\mathbf{W} \mathbf{B} \mathbf{B}^t \mathbf{W}^t) = \diag(\mathbf{Z} \mathbf{Z}^t) \leq \mathbf{1}
\end{eqnarray}

Our second approximation is to relax $\mathbf{Z}$ to a free variable instead of constraining it to $\mathbf{Z} = \mathbf{W} \mathbf{B}$ because when $\lambda$ is small and $L > M$, this free variable will converge to $\mathbf{Z} \approx \mathbf{W} \mathbf{B}$ following the same analysis in our first approximation. Now the objective function can be written as:
\begin{eqnarray}
E = \frac{\| \mathbf{B}^t - \mathbf{Z}^t \mathbf{A}^t\|_F^2}{2} + \lambda \|\mathbf{A}^t\|_1
\end{eqnarray}
Each column of $\mathbf{Z}^t$ should have L2 norm less than or equal to $1$. We see that this objective function can also be symbolically mapped to Eq~(\ref{eq:spca:sc_objective}). Hence its parameters can be optimized by efficient Sparse Coding algorithms.

\begin{figure}
    \begin{center}
    \subfigure[Objective function during learning]{\includegraphics[width=0.49\textwidth]{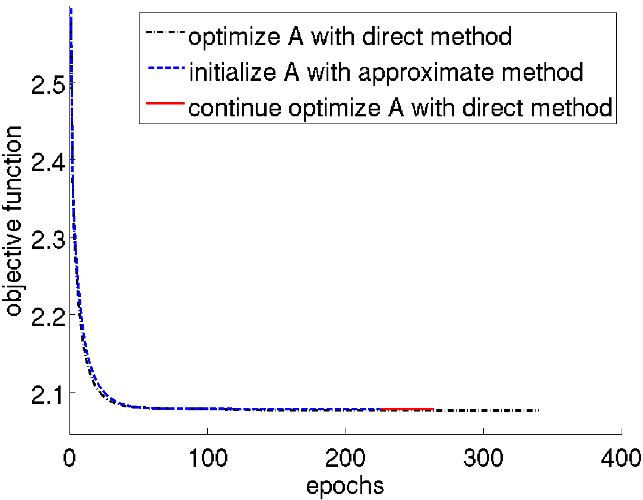}\label{fig:spca:grayscale:method:objective}}
    \subfigure[Preserved variance]{\includegraphics[width=0.49\textwidth]{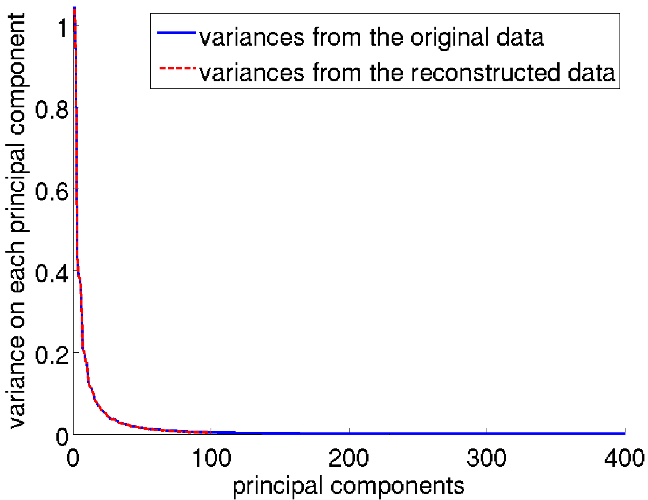}\label{fig:spca:grayscale:method:variance}}\\
    \end{center}
    \caption{Experiments on grayscale image patches. Figure~\ref{fig:spca:grayscale:method:objective} plots how the objective function changes during learning. The black dash-dotted line plots the objective function using the direct optimization method; the blue dashed line plots the objective function when we use the approximate method to find $\mathbf{A}$; the red solid line plots the objective function when we further fine tune the parameters using the direct method after initializing A with the approximate method. Figure~\ref{fig:spca:grayscale:method:variance} plots the eigenvalues of $\mathbf{C}$ from the original data (the blue solid line) versus those from the reconstructed data (the red dashed line). Our model captures $99.23\%$ of the variance that could be captured by an optimal linear model with $100$ output neurons.}
    \label{fig:spca:grayscale:method}
\end{figure}

In the grayscale image experiment, we verified the efficacy of this approximation. We tested two optimization methods: directly optimizing the objective function in Eq~(\ref{eq:spca:objective:matrix2}), or first initialize $\mathbf{A}$ by optimizing Eq~(\ref{eq:spca:final_objective}) and then optimize Eq~(\ref{eq:spca:objective:matrix2}). We implemented the experiments in single precision on a computer server with Intel Core i7 processors. When we reduce the dimensionality from $400$ to $100$ and set $\lambda=0.004$, it takes $37$ minutes to directly optimize Eq~(\ref{eq:spca:objective:matrix2}). On the other hand, it takes less than $10$ seconds to initialize $\mathbf{A}$ by optimizing Eq~(\ref{eq:spca:final_objective}) and another $40$ seconds to optimize Eq~(\ref{eq:spca:objective:matrix2}). As shown in Figure~\ref{fig:spca:grayscale:method:objective}, our approximate method efficiently minimizes the objective function. In fact, after $\mathbf{A}$ is initialized with the approximate method, further optimizing the objective function with the direct method only causes a less than $0.1\%$ change of the objective function, and on average, the weights in $\mathbf{A}$ are only changed by $0.4\%$. No visible difference in the features and filters can be observed. Because of the closeness of the approximation, in later experiments (color images, video, and sound), we used the approximation provided by Eq~(\ref{eq:spca:final_objective}) without additional optimization.

\section{URLs of Video Clips} \label{sec:spca:appendix:url}
\url{www.youtube.com/watch?v=cMIRwCNvI94} ~~~~
\url{www.youtube.com/watch?v=J7eRGHVx3p0} ~~~~
\url{www.youtube.com/watch?v=8R1g0t00vGM} ~~~~
\url{www.youtube.com/watch?v=K61FRGpvfwE} ~~~~
\url{www.youtube.com/watch?v=M-nXN5SGmhw} ~~~~
\url{www.youtube.com/watch?v=tOn2RhH36Mc} ~~~~
\url{www.youtube.com/watch?v=gc9jFmkjizQ} ~~~~
\url{www.youtube.com/watch?v=ZiW96Uci624} ~~~~
\url{www.youtube.com/watch?v=1YQrLPW5DdY} ~~~~
\url{www.youtube.com/watch?v=xKksJ3fvB1Q} ~~~~
\url{www.youtube.com/watch?v=43id_NRajDo} ~~~~
\url{www.youtube.com/watch?v=NRWehNVSAlA} ~~~~
\url{www.youtube.com/watch?v=oJ-KzdRsQC4} ~~~~
\url{www.youtube.com/watch?v=VuMRDZbrdXc} ~~~~
\url{www.youtube.com/watch?v=2rlZVtKKWnk} ~~~~
\url{www.youtube.com/watch?v=aIQB0NFcFog} ~~~~
\url{www.youtube.com/watch?v=B71T_GpA2AM} ~~~~
\url{www.youtube.com/watch?v=XB-8Grn6sRo} ~~~~
\url{www.youtube.com/watch?v=JxrIWShNPko} ~~~~
\url{www.youtube.com/watch?v=gVjqL-9Fh3E} ~~~~
\url{www.youtube.com/watch?v=yKKabd3W904} ~~~~
\url{www.youtube.com/watch?v=4ZFoqh8PQ88} ~~~~
\url{www.youtube.com/watch?v=NR3Z4p5hspI} ~~~~
\url{www.youtube.com/watch?v=RB9uzMjiYSQ} ~~~~
\url{www.youtube.com/watch?v=a7XuXi3mqYM} ~~~~
\url{www.youtube.com/watch?v=PBrStxuOJbs} ~~~~
\url{www.youtube.com/watch?v=u6ouWOGJk5E}


\end{document}